\documentclass[twocolumn,showpacs,preprintnumbers,amsmath,amssymb]{revtex4}
\input psfig.sty

\usepackage{graphicx}
\usepackage{dcolumn}
\usepackage{bm}

\begin{document}

\def\be{\begin{equation}}
\def\ee{\end{equation}}

\title{High-redshift objects and the generalized Chaplygin gas}

\author{J. S. Alcaniz} \email{alcaniz@astro.washington.edu}
\affiliation{Astronomy Department, University of Washington, Seattle,
Washington, 98195-1580, USA
}%

\author{Deepak Jain} \email{deepak@ducos.ernet.in}
\affiliation{%
Deen Dayal Upadhyaya College, University of Delhi,
Delhi - 110015, India
}%

\author{Abha Dev} \email{abha@ducos.ernet.in}
\affiliation{Department of Physics and Astrophysics, University of Delhi,
Delhi - 110007, India}

\date{\today}

\begin{abstract}

Motivated by recent developments in particle physics and cosmology, there has been growing interest in an
unified description of dark matter and dark energy scenarios. In this paper we explore observational
constraints from age estimates of high-$z$ objects on cosmological models dominated by an exotic fluid with
equation of state $p = -A/\rho^{\alpha}$ (the so-called generalized Chaplygin gas) which has the interesting
feature of interpolating between
non-relativistic matter and negative-pressure dark energy regimes. As a general result we find that, if the
age estimates of these objects are correct, they impose very restrictive limits on some of these scenarios.

\end{abstract}

\maketitle

\section{Introduction}

The question about the nature of the energy content of the Universe was always a central topic in
Cosmology. In the last
few years, however, such a discussion has become even more critical due to a convergence of observational
results
that strongly support the idea of an accelerated universe dominated by cold dark matter (CDM) and an exotic
fluid with a large negative pressure. Dark matter is inferred from galactic rotation curves which show a
general behavior that is significantly different from that one predicted by Newtonian mechanics. The most
direct evidence for the dark energy component or
``quintessence" cames
from distance measurements of type Ia supernovae (SNe Ia) which indicate that the expansion of te Universe is
speeding up, not slowing down \cite{perlmutter}. Another important evidence arises from a discrepancy
between the measurements of the cosmic microwave background (CMB) anysotropies which indicate
$\Omega_{\rm{Total}} = 1.1 \pm 0.07$ \cite{bern} and clustering estimates providing $\Omega_{m} = 0.3 \pm 01$
\cite{calb}. While the combination of these two latter results implies the existence of a smooth component of
energy that contibutes with $\simeq 2/3$ of the critical density, the SNe Ia results requires this component
to have a negative pressure which leads to a repulsive gravity.

The main distinction between these two dominant forms of energy (or matter) existent in the Universe is
manisfested through their gravitational effects. Cold dark matter agglomerates at small scales whereas the
dark energy seems to be a smooth component, a fact that is directly linked to the equation of state of both
components. Recently, the idea of a unified description for the CDM and dark energy scenarios has received
much attention \cite{uni,wat,kasu,pad}. For example, Wetterich \cite{wat} suggested that dark matter migth
consist of
quintessence lumps while Kasuya \cite{kasu} showed that spintessence-like scenarios are generally unstable to
formation of $Q$ balls which behave as pressureless matter. More recently, Padmanabhan and Choudhury 
\cite{pad}
investigated such a possibility via a string theory motivated tachyonic field.

Another interesting attempt of unification was suggested by Kamenshchik {\it et al.} \cite{kame} and developed
by Bili\'c {\it et al.} \cite{bilic}. It refers to an exotic fluid, the so-called Chaplygin gas, whose
equation of state is given by \cite{bento}
\begin{equation}
p = -A/\rho^{\alpha},
\end{equation}
with $\alpha = 1$ and $A$ a positive constant. In actual fact, the above equation for $\alpha \neq 1$
constitutes a generalization of the original Chaplygin gas equation of state recently proposed in Ref.
\cite{bento}. By inserting the Eq. (1) into the energy conservation
law we find the following expression for the density of this generalized Chaplygin gas
\begin{equation} \label{limB}
\rho_{Cg} = \left[A + B\left(\frac{R_o}{R}\right)^{3(1 + \alpha)}\right]^{\frac{1}{1 + \alpha}},
\end{equation}
or, equivalently,
\begin{equation}
\rho_{Cg} = \rho_{Cg_{o}}\left[A_s + (1 - A_s)\left(\frac{R_o}{R}\right)^{3(1 + \alpha)}\right]^{\frac{1}{1 +
\alpha}},
\end{equation}
where the subscript $o$ denotes present day quantities, $R(t)$ is the cosmological scale factor, $B =
\rho_{Cg_{o}}^{1 + \alpha} - A$ is a constant and $A_s =
A/\rho_{Cg_{o}}^{1 + \alpha}$ is a quantity related with the sound speed for the Chaplygin gas today. As can
be seen
from the above equations, the Chaplygin gas interpolates between non-relativistic matter ($\rho_{Cg}(R
\rightarrow 0) \simeq \sqrt{B}/R^{3}$) and negative-pressure dark component regimes ($\rho_{Cg}(R \rightarrow
\infty)
\simeq \sqrt{A}$).

From the theoretical viewpoint, an interesting connection between the Chaplygin gas equation of state and
String theory has been identified \cite{hoppe,hoppe1} (see also \cite{jac} for a detailed review). As
explained in
these references, a Chaplygin gas-type equation of state is
associated with the parametrization invariant Nambu-Goto $d$-brane action in a $d + 2$ spacetime. In the
light-cone parametrization, such an action reduces itself to the action of a Newtonian fluid which obeys Eq.
(1) with $\alpha = 1$ so that the Chaplygin gas corresponds effectively to a gas of $d$-branes in a $d + 2$
spacetime. Moreover, the Chaplygin gas is the only gas known to admit supersymmetric generalization
\cite{jac}. From the observational viewpoint, these cosmological scenarios have interesting features
\cite{gorini} which make them in agreement with the most recent observations of SNe Ia \cite{fabris1,
avelino,ioav}, the location of the CMB peaks \cite{b11}, age estimates of globular clusters, as well as with
the current gravitational lensing data
\cite{aba}.

In this paper we discuss new observational constraints on Chaplygin gas cosmologies from age considerations
due to the existence of three recently reported old high-redshift objects, namely, the LBDS
53W091, a 3.5-Gyr-old radio galaxy at $z = 1.55$ \cite{dunlop}, the LBDS 53W069, a 4.0-Gyr-old
radio galaxy at $z = 1.43$ \cite{dunlop1} and a quasar, the APM 08279+5255 at $z = 3.91$ whose age is
estimated between 2 - 3 Gyr \cite{karm}. Two different cases will be studied: a flat scenario in which the
generalized
Chaplygin gas together with the observed baryonic content are responsible by the dynamics of the present-day
Universe [unifying dark matter-energy] (UDME)  and a flat scenario driven by non-relativistic matter plus 
the generalized Chaplygin gas (GCgCDM). For UDME scenarios we adopt in our computations $\Omega_b = 0.04$, in
accordance with the
latest measurements of the Hubble parameter \cite{friedman} and of the baryon density at nucleosynthesis
\cite{burles}. For GCgCDM models we assume $\Omega_{m} = 0.3$, as sugested by dynamical estimates on scales up
to about $2h^{-1}$ \cite{calb}. For the sake of completeness an additional analysis for the conventional case
($\alpha = 1$) is also included. The plan of this paper is as follows. In Sec. II we present the most relevant
formulas to our analysis, as
well as the main assumptions for the age-redshift test. We then proceed to discuss the constraints provided by
this test on the cosmological scenarios decribed above in Sec. III. We end this paper by summarizing the main
results in the conclusion section.

\begin{figure}
\centerline{\psfig{figure=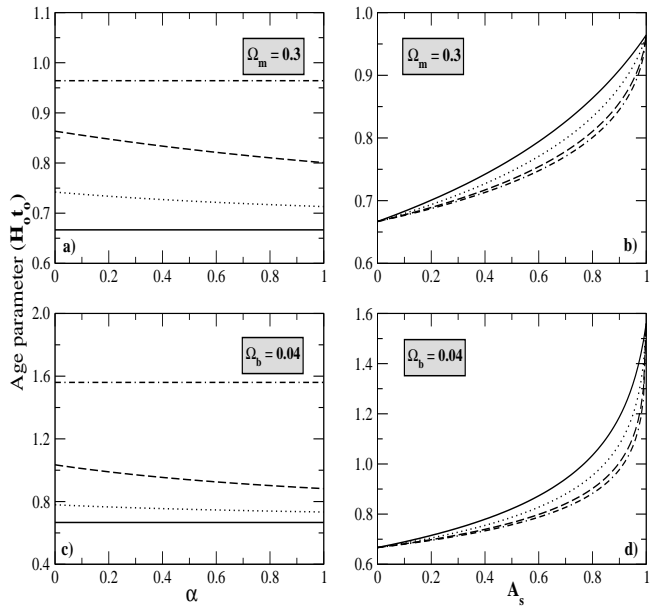,width=3.5truein,height=3.4truein,angle=-90}
\hskip 0.1in}
\caption{Dimensionless age parameter as a function of $\alpha$ and $A_s$ for UDME and GCgCDM scenarios.
In order to have a bidimensional plot we have fixed some selected values of $A_s$ in panels (a) and
(c) and $\alpha$ in panels (b) and (d). From top to botton the curves correspond to the following values
of these parameters: 1.0, 0.8, 0.4 and 0.0.}
\end{figure}

\section{Age-redshift Test}

The general Friedmann's equation for the kind of models we are considering is
\begin{eqnarray}
(\frac{{\dot R}}{R})^{2} & = &  H_o^{2} \{\Omega_{j}(\frac{R_o}{R})^{3} + \\ \nonumber & & + (1 -
\Omega_{j})[A_s + (1 - A_s)
(\frac{R_o}{R})^{3(\alpha + 1)}]^{\frac{1}{\alpha + 1}}\},
\end{eqnarray}
where $H_o$ is the present value of the Hubble parameter and $\Omega_{j}$ stands for the baryonic matter
density parameter ($j = b$) in UDME scenarios and the baryonic + dark matter density parameter ($j = m$) in
GCgCDM models.

The age-redshift relation as a function of the observable parameters is
written as
\begin{eqnarray}  \label{age}
t_z  =  {1 \over H_o} & & \int_{0}^{(1 + z)^{-1}} {dx \over x f(\Omega_{j}, A_s, \alpha,
x)} \\ \nonumber & & = {1 \over H_o} g(\Omega_{j}, A_s, \alpha, z),
\end{eqnarray}
where $x$ is a convenient integration variable and the dimensionless function
$f(\Omega_{j}, A_s, \alpha, x)$ is given by
\begin{eqnarray}
f(\Omega_{j}, A_s, \alpha, x)  =   \sqrt{\frac{\Omega_{j}}{x^{3}} + (1 - \Omega_{j})\left[A_s + \frac{(1 -
A_s)}{x^{3(\alpha + 1)}}\right]^{\frac{1}{\alpha + 1}}}
\end{eqnarray}
The total expanding age of the Universe is obtained by taking $z = 0$ in Eq. (\ref{age}). As one may check,
for $\alpha = 1$ and $A_s = 1$ Eq. (\ref{age}) reduces to the $\Lambda$CDM case while for $\alpha = 1$ and
$A_s = 0$ the standard relation [$t_z = \frac{2}{3}H_o^{-1}(1 + z)^{-\frac{3}{2}}$] is recovered. A
recent discussion about the globular clusters constraints on the total expanding age in the
context of Chaplygin gas cosmologies can be found in \cite{aba}.

In order to constrain the cosmological parameters from the age estimates of the above mentioned high-$z$
objects we take for granted that the age of the Universe at a given redshift is bigger than or at
least equal to the age of its oldest objects. In this case, the comparison of these two quantities implies a
lower (upper) bound for $A_s$ ($\alpha$), since the predicted age of the Universe increases (decreases) for
larger values of this quantity (see Fig. 1). Note also that the age parameter $H_ot_o$ is an almost
insensitive function to the parameter $\alpha$ but that it depends strongly on variations of $A_s$. This
means that age considerations will be much more efficient to constrain the sound speed $A_s$ than the values
of the parameter $\alpha$.

To quantify the above considerations we follow \cite{jailson}
and introduce the expression
\begin{equation}
\frac{t_z}{t_g} = \frac{g(\Omega_{j}, A_s, \alpha, z)}{H_o t_g} \geq 1,
\end{equation}
where $t_g$ is the age of an arbitrary object, say, a galaxy or a quasar at a given redshift $z$ and
$g(\Omega_{j}, A_s, \alpha, z)$ is the dimensionless factor defined in Eq.
(5). For each extragalatic object, the denominator of the above equation defines a dimensionless
age parameter $T_g = H_o t_g$. In particular, the 3.5-Gyr-old galaxy (53W091) at $z = 1.55$ yields
$T_g = 3.5H_o$Gyr which, for the most recent determinations of the Hubble parameter, $H_o = 72 \pm
8$ ${\rm{km s^{-1} Mpc^{-1}}}$ \cite{friedman}, takes values on the interval $0.229 \leq T_g \leq
0.286$. It thus follows that $T_g \geq 0.229$. Therefore, for a given value of $H_o$, only models
having an expanding age bigger than this value at $z = 1.55$ will be compatible with the existence
of this object. Naturally, similar considerations may also be applied to the 4.0-Gyr-old galaxy (53W069) at $z
= 1.43$ and to the 2-Gyr-old quasar (APM 08279+5255) at $z = 3.91$ . In this case, we obtain, respectively,
$T_g \geq 0.261$ and $T_g \geq 0.131$. To assure the robustness of the limits, we have  systematically adopted
in our computations the minimal value of the Hubble parameter, i.e., $H_o = 64$ ${\rm{km s^{-1} Mpc^{-1}}}$,
as well
as the underestimated age of the objects. In other words, it means that  conservative bounds are always
favored in the estimates presented here (see \cite{jailson} for details).

\begin{figure}
\centerline{\psfig{figure=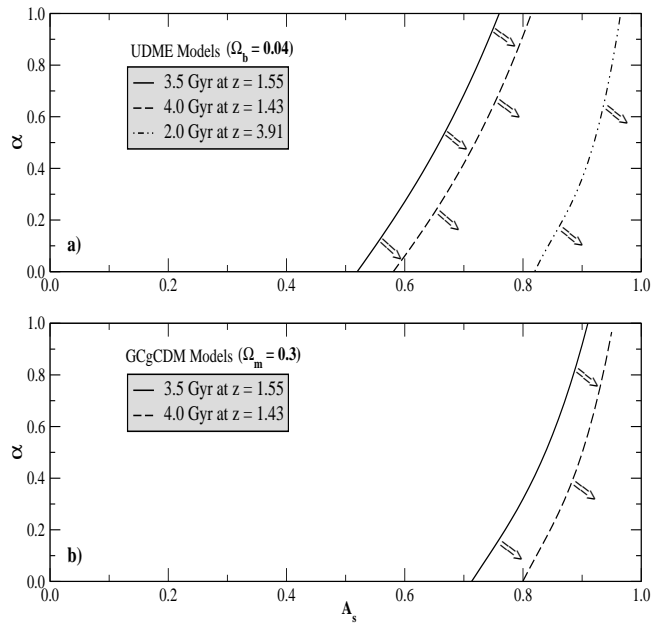,width=3.5truein,height=3.5truein,angle=-90}
\hskip 0.1in}
\caption{a - Contours of fixed age parameter $H_ot_z$ for the three high-$z$ objects discussed in the text for 
UDME models. The contours are obtained for the minimal values of $T_g$. For each contour
the arrows point to the allowed parameter space. b - The same as in panel (a) for GCgCDM scenarios.}
\end{figure}

\section{discussion}

Figures 2a and 2b show the parameter space $A_s - \alpha$ for a fixed value of the dimensionless age
parameter $H_ot_z$ for UDME and GCgCDM scenarios, respectively. For a given object, each contour represents
the minimal value
of its age parameter at the respective redshift with the arrows indicating the available parameter space
allowed by
each object. As discussed earlier, the main constraints from this kind of cosmological test are on the value
of the
parameter $A_s$ (see Fig. 1). Note also that the allowed range for this parameter is reasonably narrow. For
example, for UDME scenarios the age-redshift relation for the LBDS 53W091 and LBDS 53W069 requires,
respectively,  $A_s \geq 0.52$
and $A_s \geq 0.58$ while the same analysis for GCgCDM models provides $A_s \geq 0.72$ and $A_s \geq
0.80$. As physically expected, the limits from age considerations are much more restrictive for GCgCDM models
than for UDME scenarios. It happens because the larger the contribution of non-relativistic matter
($\Omega_j$) the
smaller the predicted age of the Universe at a given redshift and, as a consequence, the larger the value of
the parameter $A_s$ that is required in order to fit the observational data. The most restrictive bounds on
$A_s$ are provided by the quasar APM 08279+5255 at $z = 3.91$ whose age is estimated to be $\geq 2.0$ Gyr
\cite{karm}. In this case, we find $A_s \geq 0.81$ for UDME models. Our analysis also reveals that GCgCDM
scenarios with $\Omega_m = 0.3$ are not compatible with the existence of this quasar once the predicted age of
the Universe at $z = 3.91$ is smaller than the underestimated age for this object. The maximum age predicted
by this model at this redshift is 1.7
Gyr ($H_o = 64$ ${\rm{km s^{-1} Mpc^{-1}}}$) for values of $\alpha = 0$ and $A_s =1$ (the point of maximum
age; see Fig. 1). By inverting the
analysis,
i.e., by fixing the values of $\alpha$ and $A_s$, it is also possible to infer the maximum allowed value of
the matter density parameter in order to make GCgCDM models compatible with the existence of this particular
object. For $\alpha = 0$ and $A_s = 1$, we find $\Omega_m \leq 0.21$.
In other words, it means that if the age
estimates for the quasar APM 08279+5255 are correct there is an "age crisis" in the context of GCgCDM models
for
values of the matter density parameter $\Omega_m \geq 0.21$. We still recall, in line with the arguments
presented in \cite{karm},
that recent x-ray observations show an Fe/O ratio for this object that is compatible with an age of 3 Gyr. In
this
case, GCgCDM models are compatible with the existence of such a object only for values of $\Omega_m < 0.1$.
The restrictive bounds imposed by the age estimates of the quasar APM 08279+5255  on $\Lambda$CDM models,
quintessence scenarios with a equation of state $p = \omega \rho$ ($-1 \leq \omega < 0$), as well as on the
first epoch of quasar formation can be found in \cite{jnew}. 

\begin{figure}
\centerline{\psfig{figure=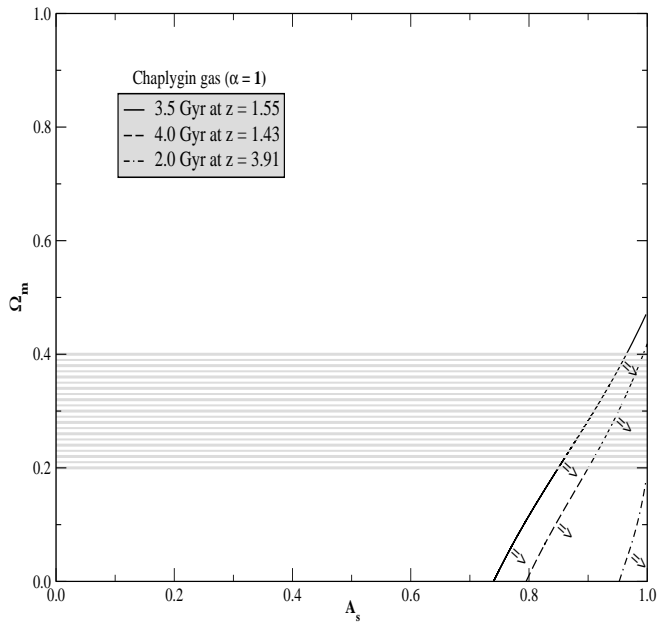,width=3.5truein,height=3.5truein,angle=-90}
\hskip 0.1in}
\caption{$A_s - \Omega_m$ plane allowed by the age estimates of the high-$z$ objects in the framework of
Chaplygin gas cosmologies ($\alpha = 1$). The shadowed region corresponds to the observed interval of the
matter density parameter $\Omega_m = 0.3 \pm 0.1$ \cite{calb}. Arrows delimit the available parameter space.
The curves are defined by the underestimated values of $t_g$ and the observed lower limit of $H_o$.}
\end{figure}

In Fig. 3 we show the $A_s-\Omega_m$ diagram allowed by the age estimates of the above mentioned objects for
the specific case in which $\alpha = 1$ (Chaplygin gas cosmologies). As in Fig. 2, the arrows indicate the
available parameter space allowed by each object. The shadowed horizontal region corresponds to the observed
interval $\Omega_m = 0.2 - 0.4$ \cite{calb} which now is used to fix the lower bounds to $A_s$. By considering 
this  interval, the LBDS 53W091 and LBDS 53W069 provides, respectively, $A_s \geq 0.85$, $A_s \geq 0.96$ and
$A_s \geq 0.90$ and $A_s \geq 0.99$. These values are even more restrictive than those obtained in the
previous analyses because the predicted age of the Universe is smaller for larger values of $\alpha$. Such
limits also provide a minimal total age of the Universe of the order of 13 Gyr. Finally, as
expected from previous analyses, the quasar APM 08279+5255 provides the most restrictive bounds on these
cosmologies. In reality, its existence
is not compatible with Chaplygin gas cosmologies ($\alpha =1$) unless the matter density parameter is $\leq
0.17$. Such a result may be used to reinforce the idea of dark matter-energy unification once UDME models are
not only compatible with the existence of these high-$z$ objects (and, as a consequence, with general age
considerations) but also provide the best fit for the SNe data \cite{fabris1}. The main results of the present
paper are summarized in Table I.

\begin{table}
\caption{Limits to $A_s$}
\begin{ruledtabular}
\begin{tabular}{lcr}
Object & UDME & GCgCDM \\
\hline \hline \\
LBDS 53W091 & $A_s \geq 0.52$ & $A_s \geq 0.58$\\
LBDS 53W069 & $A_s \geq 0.72$ &  $A_s \geq 0.80$\\
APM 08279+5255 & $A_s \geq 0.81$ & $\sim$\footnote{The entire range is imcompatible}\\
\hline \hline \\
& Chaplygin gas ($\alpha = 1$) & \\
\hline \hline \\
 & $\Omega_m = 0.2$ & $\Omega_m = 0.4$ \\
\hline \hline \\
LBDS 53W091 & $A_s \geq 0.85$ & $A_s \geq 0.96$ \\
LBDS 53W069 & $A_s \geq 0.90$ & $A_s \geq 0.99$ \\
APM 08279+5255 & $\sim$ & $\sim$ \\
\end{tabular}
\end{ruledtabular}
\end{table}

\section{conclusion}

We have investigated new observational constraints from age estimates of high-$z$ objects on generalized
Chaplygin gas cosmologies. Two diferent cases have been analysed, namely, UDME scenarios in which the dynamics
of the present day Universe is completely determined by the generalized Chaplygin gas and the observed
baryonic content ($\Omega_b = 0.04$) and GCgCDM models in which the generalized Chaplygin gas plays the role
of dark energy only and is responsable by the dynamics of the Universe together with the dark matter
($\Omega_m = 0.3$). The former kind of cosmological scenarios is inspired by the ideas of dark matter-energy
unification while the latter follows the conventional ``quintessence" program. By considering the age
estimates of the radio galaxies LBDS 53W091, LBDS 53W069 and of the quasar APM 08279+5255, we have derived
very restrictive constraints on the free parameters of these models (see Table I). In particular, we have
found that, similarly to models with a relic cosmological constant, there is no ``age of the Universe problem"
in the
context of UDME scenarios while GCgCDM models are imcompatible with the age estimates of the quasar APM
08279+5255
for values of $\Omega \geq 0.21$. Such result may be understood as a backup to the idea of dark matter-energy
unification once UDME models also provide the best fit for SNe Ia data \cite{fabris1}. However, we emphasize
that only with new and more precise set of observations will be possible to show whether or not this class of
models constitutes a viable possibility of unification for the dark matter and dark energy scenarios.

\begin{acknowledgments}
JSA is supported by the Conselho Nacional de Desenvolvimento Cient\'{\i}fico e
Tecnol\'{o}gico (CNPq - Brasil) and CNPq (62.0053/01-1-PADCT III/Milenio).
\end{acknowledgments}


\end{document}